\documentclass[aps,prd,amsmath,floats,floatfix,twocolumn,
  superscriptaddress,nofootinbib,showpacs]{revtex4-2}

\usepackage[dvipsnames]{xcolor}
\usepackage{hyperref}
\hypersetup{
  colorlinks,
  citecolor={cyan!50!black},
  linkcolor={orange!60!black},
  urlcolor=violet,
}
\usepackage{orcidlink}
\usepackage{graphicx}
\usepackage{amssymb}
\usepackage{mathtools}  %
\usepackage{tensor}
\usepackage{appendix}
\usepackage{bm} %
\usepackage{commath}  %
\usepackage{physics}
\usepackage{bbm} %
\usepackage{microtype}
\usepackage{enumitem}
\usepackage{siunitx}
\usepackage{csquotes}
\usepackage[capitalise]{cleveref}
\usepackage[caption=false]{subfig}
\crefname{section}{Sec.}{Sec.}
\Crefname{section}{Section}{Sections}
\newcommand{\ccite}[1]{Ref.~\cite{#1}}
\newcommand{\ccites}[1]{Refs.~\cite{#1}}

\usepackage{tikz}
\newcommand{\tikzinput}[1]{
  \includegraphics{#1.pdf}
}

\setlist[description]{font=\normalfont\itshape}

\newcommand{\defeq}{\coloneqq}

\newcommand{\dgF}{\mathcal{F}}
\newcommand{\dgFi}[1]{\tensor{\dgF}{_{\!{#1}}^i}}
\newcommand{\dgFj}[1]{\tensor{\dgF}{_{\!{#1}}^j}}
\newcommand{\dgFA}{\mathcal{A}}
\newcommand{\dgFB}{\mathcal{B}}
\newcommand{\dgS}{\mathcal{S}}

\newcommand{\dgf}{f}

\newcommand{\bas}{\psi}
\newcommand{\lagr}{\ell}
\newcommand{\jac}{\mathrm{J}}
\newcommand{\invjac}{{(\jac^{-1})}}
\newcommand{\surf}[1]{{#1}^\Sigma}
\newcommand{\atp}[1]{\left.#1\right|}

\newcommand{\dgM}{M}
\newcommand{\dgD}{D}
\newcommand{\dgMD}{M\!\!D}

\newcommand{\dgML}{M\!\!L}
\newcommand{\dgMLD}{M\!\!L\!D}

\newcommand{\ex}{\mathrm{ex}}
\newcommand{\bg}{\mathrm{bg}}

\newcommand{\spec}{\texttt{SpEC}}
\newcommand{\spectre}{\texttt{SpECTRE}}

\begin{document}

\title{Discontinuous Galerkin scheme for elliptic equations\\on extremely stretched grids}

\newcommand{\caltech}{\affiliation{Theoretical Astrophysics, Walter Burke
Institute for Theoretical Physics, California Institute of Technology, Pasadena,
California 91125, USA}}

\author{Nils L. Vu\,\orcidlink{0000-0002-5767-3949}} \email{nilsvu@caltech.edu} \caltech

\date{\today}

\begin{abstract}
  Discontinuous Galerkin (DG) methods for solving elliptic equations are gaining
  popularity in the computational physics community for their high-order
  spectral convergence and their potential for parallelization on computing
  clusters.
  However, problems in numerical relativity with extremely stretched grids, such
  as initial data problems for binary black holes that impose boundary
  conditions at large distances from the black holes, have proven challenging
  for DG methods.
  To alleviate this problem we have developed a primal DG scheme that is
  generically applicable to a large class of elliptic equations, including
  problems on curved and extremely stretched grids. The DG scheme accommodates
  two widely used initial data formulations in numerical relativity, namely the
  puncture formulation and the extended conformal thin-sandwich (XCTS)
  formulation.
  We find that our DG scheme is able to stretch the grid by a factor of $\sim
  10^9$ and hence allows to impose boundary conditions at large distances. The
  scheme converges exponentially with resolution both for the smooth XCTS
  problem and for the nonsmooth puncture problem.
  With this method we are able to generate high-quality initial data for binary
  black hole problems using a parallelizable DG scheme. The code is publicly
  available in the open-source \spectre{} numerical relativity code.
\end{abstract}

\maketitle

\section{Introduction}\label{sec:intro}

Discontinuous Galerkin (DG) methods are a class of spectral finite-element
numerical methods for solving partial differential equations. Although
most prevalently applied to hyperbolic conservation laws, DG methods can also be
advantageous for elliptic equations~\cite{Arnold2002, Wheeler1978-tj}. Their
spectral elements are able to retain exponential convergence
rates~\cite{Schoetzau2014}, and the domain decomposition allows for efficient
parallelization on large computing
clusters~\cite{Vincent2019qpd,dgscheme1,ellsolver}. Together, these properties
make DG methods attractive for computationally demanding elliptic problems in
computational physics, such as binary black hole initial data in numerical
relativity~\cite{Cook2000lrr, Pfeiffer2004-oo}. High-accuracy initial data is
particularly important to underpin ongoing efforts in the numerical relativity
community to develop a next generation of massively parallel simulation
codes~\cite{spectre,CarpetX,Fernando:2018mov,Tichy:2022hpa,Daszuta:2021ecf,Reinarz:2019:ExaHyPE,Zhu:2024utz},
specifically to simulate binary black hole mergers with sufficient accuracy for
the next generation of gravitational wave
observatories~\cite{Purrer:2019jcp,Ferguson:2020xnm,Jan:2023raq}.

However, elliptic problems in the field of numerical relativity often involve
extremely stretched grids to impose boundary conditions at very large distances,
which can be challenging for DG methods. Specifically, initial data problems in
general relativity with black holes or neutron stars are typically solved on
spherical domains centered on the objects. Since the solution at the outer
boundary of the sphere is unknown but typically falls off as $1/r$ with
coordinate distance from the center, the boundary condition is set to empty flat
spacetime at a large but finite radius $R$ (``approximate asymptotic
flatness'')~\cite{Pfeiffer2003-mt}. This approach incurs an error of
$\mathcal{O}(1/R)$ on the solution, which is reduced below numerical accuracy by
pushing the outer boundary far enough out, often to $R \sim 10^9$. To resolve
this large computational volume numerically, radial grid points are also
distributed on a $1/r$ scale. This way, a large spherical shell can be resolved
by just a few spectral elements that are extremely stretched. \Cref{fig:wedge}
below illustrates such an extremely stretched spectral element. While this
strategy has worked well for specialized spectral numerical schemes in the
past~\cite{Pfeiffer2003-mt}, generic DG schemes struggle with the extremely
large Jacobian factors that arise from the grid stretching. Specifically, we
found that the DG scheme developed in \ccite{dgscheme1} fails to converge on the
extremely stretched grids used in numerical relativity. In this article we
construct a primal DG scheme for generic elliptic equations that is capable of
handling such extreme grid stretching. The scheme is implemented in the
open-source numerical relativity code \spectre{}~\cite{spectre}.

To further alleviate the strain that the extreme grid stretching can pose on the
numerical scheme, we propose new Robin-type boundary conditions at the outer
boundary. These boundary conditions incur an error of only $\mathcal{O}(1/R^2)$,
so the outer boundary can be placed at a smaller radius, $R\sim 10^5$, while
still maintaining the same level of error. To this end, we develop new
Robin-type boundary conditions for the extended conformal thin sandwich (XCTS)
formulation of the Einstein constraint equations.

Alternative approaches to the extreme grid stretching have been proposed on the
level of the elliptic equations themselves. For example, the
\texttt{LORENE}~\cite{Lorene, Grandclement:2006ht},
\texttt{KADATH}~\cite{Grandclement2010-wk,Papenfort:2021hod}, and
\texttt{SGRID}~\cite{Tichy:2009yr} libraries employ multiple computational
domains and a radial coordinate transformation to compactify the outermost
domain. This allows them to impose boundary conditions at spatial infinity rather
than at a finite radius, but requires special treatment of derivative operators
in the different domains to incorporate the coordinate transformations. The
\texttt{TwoPunctures} code~\cite{Ansorg:2004ds} employs a similar
compactification on a single domain, rewriting the elliptic equations in
compactified coordinates. Other possible approaches include conformal
compactification methods and hyperboloidal
slicing~\cite{Buchman:2009ew,Schinkel:2013zm,Vano-Vinuales:2023yzs}. These
methods also require special treatment of spatial infinity in the elliptic
equations and have not yet been routinely applied to construct and evolve binary
black hole initial data. With the numerical scheme developed in this article we
avoid such special transformations of the elliptic equations to deal with
asymptotic boundary conditions at spatial infinity. Instead, the equations can
simply be solved on a sufficiently large domain such that the error incurred by
the finite outer radius falls below numerical accuracy.

This article is structured as follows. \Cref{sec:dg} formulates the primal
DG scheme for generic elliptic equations. \Cref{sec:tests} demonstrates the
versatility of the DG scheme by solving a simple Poisson problem and two classic
general-relativistic initial data problems with black holes on extremely
stretched grids. We conclude in \cref{sec:conclusion}.

\section{Discontinuous Galerkin formulation}\label{sec:dg}

The discontinuous Galerkin (DG) formulation developed here applies to any set of
coupled, nonlinear elliptic equations that can be written in first-order flux
form,
\begin{align}\label{eq:fluxform}
  -\partial_i \dgFi{\alpha}[u,\partial u;\bm{x}]
  + \dgS_\alpha[u,\partial u;\bm{x}]
  = \dgf_\alpha(\bm{x})
  \text{,}
\end{align}
where $\dgFi{\alpha}$ are the fluxes, $\dgS_\alpha$ are the sources, and
$\dgf_\alpha$ are the fixed sources. The index~$\alpha$ enumerates the set of
equations and variables. The fluxes and sources are functionals of the
variables~$u_\alpha(\bm{x})$ and their first derivatives. The
fluxes~$\dgFi{\alpha}$ must be linear in the variables and their derivatives,
which means they can be written in the form
\begin{equation}\label{eq:fluxdecomp}
  \dgFi{\alpha} = \dgFA_{\alpha\beta}^{ij} \partial_j u_\beta + \dgFB_{\alpha\beta}^i u_\beta
  \text{,}
\end{equation}
where $\dgFA_{\alpha\beta}^{ij}$ must be positive definite to ensure ellipticity
of the equations. The sources~$\dgS_\alpha$ can be nonlinear. The fixed
sources~$\dgf_\alpha(\bm{x})$ are independent of the variables.
\Cref{eq:fluxform} closely resembles
formulations of hyperbolic conservation laws but allows the
fluxes~$\dgFi{\alpha}$ to be higher-rank tensor fields. It is a simplification
of the first-order flux form defined in \ccite{dgscheme1} that arises by
selecting the gradients of the variables as first-order auxiliary variables.

For example, a Poisson equation for the variable $u(\bm{x})$ on a curved
manifold with metric $g_{ij}(\bm{x})$ in Cartesian coordinates is
\begin{equation}\label{eq:poisson}
  -g^{ij} \nabla_i \nabla_j u(\bm{x}) = f(\bm{x})
  \text{,}
\end{equation}
where $\nabla_i$ is the covariant derivative compatible with the metric
$g_{ij}$. It can be written in first-order flux form~\eqref{eq:fluxform} with
the fluxes and sources
\begin{align}
  \dgF^i = g^{ij} \partial_j u \text{,} \quad
  \dgS = -\Gamma^i_{ij} \dgF^j
  \text{,}
\end{align}
where $\Gamma^i_{jk} = \frac{1}{2} g^{il} \left(\partial_j g_{kl} + \partial_k
g_{jl} - \partial_l g_{jk}\right)$ are the Christoffel symbols associated with
the metric $g_{ij}$. Any Poisson equation with nonlinear sources, such as the
puncture equation solved in \cref{sec:punctures} for simple binary black hole
initial data, can be written in this form as well.

Also the extended conformal thin-sandwich (XCTS) formulation of the Einstein
constraint equations~\cite{York:1998hy,Pfeiffer:2002iy,Pfeiffer2004-oo}, that we
solve for binary black hole initial data in \cref{sec:xcts}, can be written in
first-order flux form~\eqref{eq:fluxform}. The XCTS equations are a set of five
coupled, nonlinear elliptic equations for the variables
$\{\psi,\alpha\psi,\beta^i\}$ given by
\begin{subequations}\label{eq:xcts}
\begin{align}
  \label{eq:xcts_hamiltonian}
  \bar{\nabla}^2 \psi &= \begin{aligned}[t]
    &\frac{1}{8}\psi\bar{R} + \frac{1}{12}\psi^5 K^2 \\
    &-\frac{1}{8}\psi^{-7}\bar{A}_{ij}\bar{A}^{ij} - 2\pi\psi^5\rho
  \end{aligned}
  \\
  \label{eq:xcts_lapse}
  \bar{\nabla}^2\left(\alpha\psi\right) &=
  \begin{aligned}[t]
    &\alpha\psi \bigg(\frac{7}{8}\psi^{-8}\bar{A}_{ij}\bar{A}^{ij} + \frac{5}{12}\psi^4 K^2 + \frac{1}{8}\bar{R} \\
    &+ 2\pi\psi^4\left(\rho + 2S\right)\bigg) - \psi^5\partial_t K + \psi^5\beta^i\bar{\nabla}_i K
  \end{aligned}
  \\
  \label{eq:xcts_momentum}
  \bar{\nabla}_i(\bar{L}\beta)^{ij} &= \begin{aligned}[t]
    &(\bar{L}\beta)^{ij}\bar{\nabla}_i \ln(\bar{\alpha}) + \bar{\alpha}\bar{\nabla}_i\left(\bar{\alpha}^{-1}\bar{u}^{ij}\right) \\
    &+ \frac{4}{3}\bar{\alpha}\psi^6\bar{\nabla}^j K + 16\pi\bar{\alpha}\psi^{10}S^j
  \end{aligned}
\end{align}
\end{subequations}
with $\bar{A}^{ij} = \frac{1}{2\bar{\alpha}}\left((\bar{L}\beta)^{ij} -
\bar{u}^{ij}\right)$ and $\bar{\alpha} = \alpha \psi^{-6}$. The XCTS equations
are formulated on a background geometry defined by the conformal
metric~$\bar{\gamma}_{ij}$, which also defines the covariant
derivative~$\bar{\nabla}$, the Ricci scalar~$\bar{R}$, and the longitudinal
operator
\begin{equation}\label{eq:longitudinal_op}
  \left(\bar{L}\beta\right)^{ij} = \bar{\nabla}^i\beta^j + \bar{\nabla}^j\beta^i
  - \frac{2}{3}\bar{\gamma}^{ij}\bar{\nabla}_k\beta^k
  \text{.}
\end{equation}
The conformal metric is chosen in advance alongside the trace of the extrinsic
curvature~$K$, their respective time derivatives~$\bar{u}_{ij}$ and $\partial_t
K$, and the matter sources $\rho$, $S$, and~$S^i$. \Cref{sec:xcts_fluxes}
lists the fluxes and sources to formulate the XCTS equations in first-order flux
form~\eqref{eq:fluxform}.

Therefore, a generic implementation of the DG discretization for the first-order
flux form~\eqref{eq:fluxform} encompasses a large class of elliptic equations,
including the puncture and XCTS initial data formulations for black holes. It
also encompasses other (nonlinear) Poisson-type equations, as well as linear
elasticity~\cite{thermalnoise}. The versatility of this generic formulation
lowers the barrier of entry for using the DG method in our code, since new
equations can be added by implementing their fluxes and sources without having
to learn the details of the DG discretization scheme. In the following we will
derive the DG discretization scheme without reference to any specific elliptic
equations.

\subsection{Domain decomposition}\label{sec:dg-domain}

\begin{figure}
  \centering
  \tikzinput{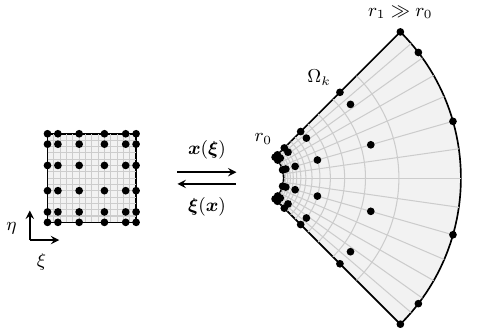}
  \caption{
    \label{fig:wedge}
    Geometry of an extremely stretched wedge-shaped element $\Omega_k$ in two
    dimensions. The coordinate transformation~$\bm{x}(\bm{\xi})$ deforms a
    reference cube~$\bm{\xi}\in[-1,1]^2$ to a wedge with inner radius $r_0$ and
    outer radius $r_1 \gg r_0$. Coordinates are mapped with an $1/r$ inverse
    radial scaling, as illustrated by the light gray lines. Grid points are
    chosen here as LGL collocation points on the reference cube (black dots).}
\end{figure}

To formulate the DG scheme we decompose the computational domain $\Omega$ into
deformed cubes, called elements~$\Omega_k$, as described in more detail in
\ccites{Vincent2019qpd,dgscheme1}. In this article we focus on extremely
stretched wedge-shaped elements as pictured in \cref{fig:wedge} from which we
build spherical domains.

To deform the hexahedral elements to wedges we employ an invertible map from
\emph{logical} coordinates $\bm{\xi}=(\xi,\eta,\zeta)\in[-1,1]^3$ on the
reference cube to the coordinates $\bm{x}\in\Omega_k$ in which the elliptic
equations~\eqref{eq:fluxform} are formulated. The coordinate map also defines
the Jacobian matrix
\begin{equation}
  \jac^i_{\hat{\jmath}} \defeq \frac{\partial x^i}{\partial \xi^{\hat{\jmath}}}
\end{equation}
with determinant $\jac$ and inverse $\invjac^{\hat{\jmath}}_i=\partial\xi^{\hat{\jmath}}/\partial x^i$.
The wedge coordinate map pictured in \cref{fig:wedge} is
\begin{equation}
  \bm{x}(\bm{\xi}) = \frac{r(\xi)}{\sqrt{1 + \eta^2 + \zeta^2}} \begin{pmatrix}
    1 \\ \eta \\ \zeta \end{pmatrix}
\end{equation}
in three dimensions, where $r(\xi)$ is the radial coordinate mapping. We choose
either a linear, logarithmic, or inverse radial mapping,
\begin{subequations}
\begin{align}
  \text{linear} &\quad r(\xi) = \frac{1}{2} \left((1-\xi) r_0 + (1+\xi) r_1\right) \\
  \text{logarithmic} &\quad \ln r(\xi) = \frac{1}{2} \left((1-\xi) \ln r_0 + (1+\xi) \ln r_1\right) \\
  \text{inverse} &\quad \frac{1}{r(\xi)}
    = \frac{1}{2} \left(\frac{1 - \xi}{r_0} + \frac{1 + \xi}{r_1}\right)
  \label{eq:invrad}
  \text{.}
\end{align}
\end{subequations}
Here, $r_0$ and $r_1$ denote the inner and outer radii of the wedge.

For illustration, we can quantify how much the reference cube is stretched by
the wedge map with the inverse radial mapping, \cref{eq:invrad}. First, note
that the midpoint of the reference cube, $\xi=0$, is mapped to only
$r_\text{mid} = 2 r_0$ in the limit $r_1 \gg r_0$. Second, the Jacobian
determinant at the outer radius is $\jac \propto (r_1)^4 / r_0$ in the limit $r_1
\gg r_0$. This means that for $r_0=10$ and $r_1=10^9$ the midpoint is mapped to
only $r_\text{mid} \sim 20$ and the Jacobian determinant is $\jac \sim 10^{35}$.
Our DG scheme has to be able to handle such extreme stretching.

For the DG discretization we choose a regular grid of
$N_k=\prod_{\hat{\imath}=1}^3 N_{k,\hat{\imath}}$ Legendre-Gauss-Lobatto~(LGL)
or Legendre-Gauss~(LG)
collocation points $\bm{\xi}_p=(\xi_{p_1},\eta_{p_2},\zeta_{p_3})$ on the
reference cube of element~$\Omega_k$, as illustrated in \cref{fig:wedge} and
described in more detail in \ccite{dgscheme1}. Fields on the computational
domain are represented by their values at these grid points. This nodal
representation is facilitated by expanding each field~$u(\bm{x})$ on the grid as
\begin{equation}\label{eq:lagr_expansion}
  u(\bm{x}) = \sum_p^{N_k} u_p \bas_p(\bm{\xi}(\bm{x}))
  \text{,}
\end{equation}
where $u_p = u(\bm{x}(\bm\xi_p))$ are the nodal coefficients and
$\bas_p(\bm{\xi})$ are the basis functions
\begin{equation}\label{eq:lagr_basis}
  \bas_p(\bm{\xi}) = \prod_{\hat{\imath}=1}^3 \lagr_{p_{\hat{\imath}}}(\xi^{\hat{\imath}})
  \text{.}
\end{equation}
These basis functions are products of the one-dimensional Lagrange interpolating
polynomials $\lagr_{p_{\hat{\imath}}}(\xi)$ with respect to the collocation
points $\bm{\xi}_p$ on the reference cube.

The choice between LGL and LG collocation points is particularly relevant on
extremely stretched grids. On LGL grids, points are placed on element
boundaries, as illustrated in \cref{fig:wedge}. On LG grids, points are only
placed in the interior of the element.
For illustration, in the previous example of a wedge with an inverse radial
mapping extending from $r_0=10$ to $r_1=10^9$, the outermost radial grid point
for LGL collocation is at the outer radius of $r=10^9$, but for LG collocation
with six radial grid points it is only at $r\approx 300$.
This can make dense matrix operations over the grid points within an element
numerically better behaved on LG grids because functions that fall off as $1/r$
vary over fewer orders of magnitude when evaluated at the LG grid points.
However, the lack of grid points at the element boundaries means that boundary
conditions must be projected into the interior of the element, which can
introduce additional errors. In this article we find that both LGL and LG
collocation points can be used successfully on extremely stretched grids.

\subsection{DG residuals}\label{sec:dg-residuals}

To derive the primal DG formulation for the elliptic equations, we project the
flux form, \cref{eq:fluxform}, on the set of basis functions~$\psi_p$ to obtain
the DG residuals
\begin{equation}\label{eq:dgres}
  -(\bas_p, \partial_i \dgF^i)_{\Omega_k} + (\bas_p, \dgS)_{\Omega_k} = (\bas_p, \dgf)_{\Omega_k}
  \text{,}
\end{equation}
where the inner product is defined as
\begin{subequations}\label{eq:dgproj}
\begin{align}
  (\phi, \pi)_{\Omega_k} \defeq{}& \int_{\Omega_k} \phi(\bm{x}) \pi(\bm{x}) \dd{^3x} \\
  ={}& \int_{[-1,1]^3} \phi(\bm{x}(\bm{\xi})) \pi(\bm{x}(\bm{\xi})) \, \jac \dd{^3\xi}
  \text{.}
\end{align}
\end{subequations}
We suppress the index~$\alpha$ that enumerates equations and variables in this
section. A partial integration of \cref{eq:dgres} gives the weak form of the DG
residuals,
\begin{equation}\label{eq:dgres2}
  (\partial_i \bas_p, \dgF^i)_{\Omega_k}
  - (\bas_p, (n_i \dgF^i)^*)_{\partial \Omega_k}
  + (\bas_p, \dgS)_{\Omega_k} = (\bas_p, \dgf)_{\Omega_k}
  \text{,}
\end{equation}
where the inner product on the element boundary is defined as
\begin{subequations}
\begin{align}
  (\phi, \pi)_{\partial\Omega_k} \defeq{}& \int_{\partial\Omega_k} \phi(\bm{x}) \pi(\bm{x}) \dd{^2x} \\
  ={}& \int_{[-1,1]^2} \phi(\bm{x}(\bm{\xi})) \pi(\bm{x}(\bm{\xi})) \, \surf{\jac} \dd{^2\xi}
\end{align}
\end{subequations}
with the surface Jacobian $\surf{\jac}$. We have also introduced the numerical
fluxes $(n_i \dgF^i)^*$ on element boundaries in \cref{eq:dgres2}, which couple
the DG residuals on neighboring elements as discussed in \cref{sec:numflux}.
Compared to \ccite{dgscheme1} we have simplified the Galerkin projections to be
performed over coordinate volume rather than proper volume because we found no
significant difference in performance between the two.

Now, to discretize the auxiliary first-order equation for the fluxes~$\dgF^i$ we
employ a primal formulation rather than the Schur-complement approach used in
\ccite{dgscheme1}. This is because the Schur-complement approach, while simpler
to implement, fails to handle extremely stretched grids. The two approaches
differ only in the boundary correction term arising from the auxiliary equation
for the fluxes~$\dgF^i$. We project the linear decomposition of the fluxes,
\cref{eq:fluxdecomp}, on the derivatives of the basis functions to find an
expression for the first term in \cref{eq:dgres2}, and integrate by parts, to
obtain
\begin{subequations}
\begin{align}
  (\partial_i \bas_p, \dgF^i)_{\Omega_k}
  &= (\partial_i \bas_p, \dgFA^{ij} \partial_j u)_{\Omega_k}
  + (\partial_i \bas_p, \dgFB^{i} u)_{\Omega_k} \\
  &= \begin{aligned}[t]
    &-(\partial_j \dgFA^{ij} \partial_i \bas_p, u)_{\Omega_k}
    + (\partial_i \bas_p, \dgFB^{i} u)_{\Omega_k} \\
    &+ (\partial_i \bas_p, \dgFA^{ij} n_j u^*)_{\partial \Omega_k}
    \text{,}
  \end{aligned}
\end{align}
\end{subequations}
where $u^*$ is the auxiliary numerical flux discussed in \cref{sec:numflux}.
With another partial integration we find
\begin{subequations}
\begin{align}
  (\partial_i \bas_p, \dgF^i)_{\Omega_k}
  &= \begin{aligned}[t]
    & (\partial_i \bas_p, \dgFA^{ij} \partial_j u)_{\Omega_k}
    + (\partial_i \bas_p, \dgFB^{i} u)_{\Omega_k} \\
    &+ \big(\partial_i \bas_p, \dgFA^{ij} n_j (u^* - u)\big)_{\partial \Omega_k}
  \end{aligned} \\
  &= (\partial_i \bas_p, \dgF^{i})_{\Omega_k}
  + \big(\partial_i \bas_p, \dgFA^{ij} n_j (u^* - u)\big)_{\partial \Omega_k}
  \text{.}
\end{align}
\end{subequations}
Inserting this result into \cref{eq:dgres2} we find the primal DG residuals in
weak form,
\begin{align}\label{eq:dgres_weak}
\begin{split}
  (\partial_i \bas_p, \dgF^{i})_{\Omega_k}
  &+ \big(\partial_i \bas_p, \dgFA^{ij} n_j (u^* - u)\big)_{\partial \Omega_k} \\
  &- (\bas_p, (n_i \dgF^i)^*)_{\partial \Omega_k}
  + (\bas_p, \dgS)_{\Omega_k} = (\bas_p, \dgf)_{\Omega_k}
  \text{,}
\end{split}
\end{align}
where $u^*$ and $(n_i \dgF^i)^*$ are the numerical fluxes discussed in the next
section.
With a final partial integration we find the primal DG residuals in strong form,
\begin{align}\label{eq:dgres_strong}
\begin{split}
  -(\bas_p, \partial_i \dgF^{i})_{\Omega_k}
  &+ \big(\partial_i \bas_p, \dgFA^{ij} n_j (u^* - u)\big)_{\partial \Omega_k} \\
  &- (\bas_p, (n_i \dgF^i)^* - n_i \dgF^i)_{\partial \Omega_k} \\
  &+ (\bas_p, \dgS)_{\Omega_k} = (\bas_p, \dgf)_{\Omega_k}
  \text{.}
\end{split}
\end{align}
Written in terms of discrete matrices over the element's grid points,
as detailed in \ccite{dgscheme1}, the primal DG residuals in weak form are
\begin{align}\label{eq:dgres_weak_matrix}
\begin{split}
  \dgMD^T_i \cdot \dgF^i
  &+ \dgMLD^T_i \cdot \dgFA^{ij} n_j (u^* - u) \\
  &- \dgML \cdot (n_i \dgF^i)^*
  + \dgM \cdot \dgS = \dgM \cdot \dgf
  \text{,}
\end{split}
\end{align}
and in strong form they are
\begin{align}\label{eq:dgres_strong_matrix}
\begin{split}
  -\dgMD_i \cdot \dgF^i
  &+ \dgMLD^T_i \cdot \dgFA^{ij} n_j (u^* - u) \\
  &- \dgML \cdot \big((n_i \dgF^i)^* - n_i \dgF^i\big)
  + \dgM \cdot \dgS = \dgM \cdot \dgf
  \text{,}
\end{split}
\end{align}
where $\dgF^i = \dgFA^{ij} \invjac^{\hat{\jmath}}_j \dgD_{\hat{\jmath}} \cdot u
+ \dgFB^i u$. The symbol $\cdot$ denotes matrix-vector multiplication over grid
points. We have defined here the mass matrix,
\begin{align}
  &\dgM_{pq} = \delta_{pq} \atp{\jac}_p \prod_{\hat{\jmath}=1}^3 w_{p_{\hat{\jmath}}}
  \text{,}
\end{align}
the weak and strong stiffness matrices in the volume of the element,
\begin{align}
  &\dgMD^T_{i,pq} = \dgD^T_{\hat{\imath},pq} \atp{\jac}_q \atp{\invjac^{\hat{\imath}}_i}_q \prod_{\hat{\jmath}=1}^3 w_{q_{\hat{\jmath}}}
  \text{,} \\
  \label{eq:strongstiff}
  &\dgMD_{i,pq} = \dgD_{\hat{\imath},pq} \atp{\jac}_q \atp{\invjac^{\hat{\imath}}_i}_q \prod_{\hat{\jmath}=1}^3 w_{p_{\hat{\jmath}}}
  \text{,}
\end{align}
as well as the lifting matrices from the element boundary to the volume of the element,
\begin{align}
  &\dgML_{p\bar{q}} = I_{p\bar{q}} \atp{\surf{\jac}}_{\bar{q}} \prod_{\hat{\jmath}=1}^2 w_{\bar{q}_{\hat{\jmath}}}
  \text{,} \\
  &\dgMLD^T_{i,p\bar{q}} = \dgD^T_{\hat{\imath},pq} I_{q\bar{q}} \atp{\surf{\jac}}_{\bar{q}} \atp{\invjac^{\hat{\imath}}_i}_{\bar{q}} \prod_{\hat{\jmath}=1}^2 w_{\bar{q}_{\hat{\jmath}}}
  \text{.}
\end{align}
To obtain these matrices we have evaluated the integrals in \cref{eq:dgres_weak}
over the reference cube using Gauss-Lobatto quadrature (on LGL grids) or Gauss
quadrature (on LG grids)~\cite{dgscheme1}.
The coefficients $w_{p_{\hat{\jmath}}}$ are the quadrature weights
in dimension $\hat{\jmath}$ of the reference cube.
Indices with a bar, i.e., $\bar{q}$, only enumerate grid points on the boundary
of the element. The boundary points are ``lifted'' to the volume of the element
with the interpolation matrices
\begin{equation}
  I_{p\bar{q}}=\lagr_{p_{\hat{\imath}}}(\pm 1)\prod_{\hat{\jmath}\neq \hat{\imath}}\delta_{p_{\hat{\jmath}} \bar{q}_{\hat{\jmath}}}
\end{equation}
for a face in dimension $\hat{\imath}$ of the reference cube. The interpolation
matrices reduce to $I_{p\bar{q}}=\delta_{p\bar{q}}$ for LGL grids, since
boundary points $\xi^{\hat{\imath}}=\pm 1$ are also grid points.
The logical differentiation matrix is defined as
\begin{equation}\label{eq:diffmat}
\dgD_{\hat{\imath},pq}=\lagr^\prime_{q_{\hat{\imath}}}(\xi_{p_{\hat{\imath}}})\prod_{\hat{\jmath}\neq \hat{\imath}}\delta_{p_{\hat{\jmath}} q_{\hat{\jmath}}}
\text{.}
\end{equation}
It takes the derivative along dimension $\hat{\imath}$ on the reference cube.

For the weak form, \cref{eq:dgres_weak_matrix}, only the $\dgMLD^T_i$ term is
different from the Schur-complement approach used in \ccite{dgscheme1},
Eq.~(30b). It involves a lifting operation with the weak differentiation matrix
and only Jacobians evaluated on the element faces.
For the strong form, \cref{eq:dgres_strong_matrix}, the volume divergence term
is different as well, specifically \cref{eq:strongstiff}. It has been shown in
\ccite{Teukolsky2016-ja} that the strong and weak forms are equivalent when
using the metric identities to move the Jacobian into the strong divergence,
\begin{align}
  \partial_i \dgF^i
  \label{eq:nometricid}
  &= \invjac^{\hat{\imath}}_i \partial_{\hat{\imath}} \dgF^i \\
  \label{eq:metricid}
  &= \frac{1}{\jac} \partial_{\hat{\imath}} \, \jac \, \invjac^{\hat{\imath}}_i \dgF^i
  \text{.}
\end{align}
We call \cref{eq:metricid} the ``strong logical'' form of the DG residuals when
used in \cref{eq:dgres_strong} because it involves a divergence in logical
coordinates. This form leads to the strong logical stiffness matrix,
\cref{eq:strongstiff}, which differs from the strong stiffness matrix in
\ccite{dgscheme1}, Eq.~(41), by the ordering of Jacobian operations. Both the
weak form and the strong logical form work equally well on extremely stretched
grids, as they are numerically equivalent. However, the strong form without
applying the metric identities, \cref{eq:nometricid}, as used in
\ccite{dgscheme1}, does not work on extremely stretched grids. We suspect that
the mixing of derivatives through the Jacobian in \cref{eq:nometricid} is
numerically ill behaved on such grids.

\subsection{Generalized internal penalty flux}\label{sec:numflux}

For the numerical fluxes in the DG residuals~\eqref{eq:dgres_weak} we choose the
generalized internal penalty flux introduced in \ccite{dgscheme1}. Using the
simplified first-order flux form, \cref{eq:fluxform}, the generalized internal
penalty flux is
\begin{subequations}\label{eq:numflux}
\begin{align}
u^* &= \frac{1}{2} \left(u^\mathrm{int} + u^\mathrm{ext}\right) \\
(n_i \dgF^i)^* &= \begin{aligned}[t]
  &\frac{1}{2} n_i \left(\dgFi{\mathrm{int}} + \dgFi{\mathrm{ext}} \right) \\
    &- \sigma \, n_i \dgFA^{ij} n_j (u^\mathrm{int} - u^\mathrm{ext})
    \text{,}
  \end{aligned}
\end{align}
\end{subequations}
where ``int'' denotes values on the interior side of a boundary shared with
another element, and ``ext'' denotes values on the exterior side that were sent
from the neighboring element. The penalty factor~$\sigma$ is
\begin{equation}
\sigma = C \, \frac{(\max(p^\mathrm{int}, p^\mathrm{ext}) + 1)^2}{\min(h^\mathrm{int},h^\mathrm{ext})}
\text{,}
\end{equation}
where $p$ is the polynomial order perpendicular to the element boundary, and
$h=2\,\jac/\surf{\jac}$ is a measure of the element size perpendicular to the
element boundary. We choose the penalty parameter $C=1.5$ in this article.

Boundary conditions are imposed through numerical fluxes as well. They are
imposed on the \emph{average} of the boundary fluxes to obtain faster
convergence~\cite{dgscheme1,HesthavenWarburton}. Therefore, for Dirichlet-type
boundary conditions $u=u_D$ on an element boundary we impose
\begin{subequations}\label{eq:dirichlet}
\begin{align}
  \frac{1}{2}(u^\mathrm{int} + u^\mathrm{ext}) &= u_D \\
  \implies \quad
  u^\mathrm{ext} &= 2 \, u_D - u^\mathrm{int}
\end{align}
\end{subequations}
and $\dgFi{\mathrm{ext}} = \dgFi{\mathrm{int}}$ in the numerical fluxes,
\cref{eq:numflux}.
For Neumann-type boundary conditions $n_i\dgF^i=(n_i\dgF^i)_\mathrm{N}$ we
instead impose
\begin{subequations}\label{eq:neumann}
\begin{align}
  \frac{1}{2}n_i(\dgFi{\mathrm{int}} + \dgFi{\mathrm{ext}})&=(n_i\dgF^i)_\mathrm{N} \\
  \implies \quad
  n_i\dgFi{\mathrm{ext}} &= 2 \, (n_i\dgF^i)_\mathrm{N} - n_i\dgFi{\mathrm{int}}
\end{align}
\end{subequations}
and $u^\mathrm{ext} = u^\mathrm{int}$ in the numerical fluxes,
\cref{eq:numflux}.

\section{Test problems}\label{sec:tests}

We test the DG scheme on three problems with extremely stretched grids. All test
problems are implemented in the open-source code \spectre{}~\cite{spectre} and
solved with the iterative multigrid-Schwarz preconditioned Newton-Krylov
elliptic solver detailed in \ccite{ellsolver}.

In addition to demonstrating that the generic DG scheme can handle extremely
stretched grids, we also demonstrate that we can obtain a measure of the
truncation error in each element and dimension of these extremely stretched
grids, which is an essential ingredient for adaptive mesh refinement (AMR). To
this end we evaluate the anisotropic relative truncation error estimate detailed
in \ccite{Szilagyi2014fna}, Eq.~(57), which has been used in the \spec{} code
for a long time~\cite{sxscatalog2019}. In logical dimension~$\hat{\imath}$ of an
element~$\Omega_k$ it is given by
\begin{equation}
  \mathcal{T}[P_{q_{\hat{\imath}}}] = \log_{10}(\max(P_1,P_2)) - \frac{\sum_{q_{\hat{\imath}}=1}^{N_{k,i}} \log_{10}(P_{q_{\hat{\imath}}}) \, W_{q_{\hat{\imath}}}}{\sum_{q_{\hat{\imath}}=1}^{N_{k,\hat{\imath}}} W_{q_{\hat{\imath}}}}
  \text{,}
\end{equation}
with the exponential weights
\begin{equation}
  W_{q_{\hat{\imath}}}=\exp(-(q_{\hat{\imath}} - N_{k,\hat{\imath}} + 1/2)^2)
  \text{.}
\end{equation}
Here, $P_{q_{\hat{\imath}}}$ are the ``power monitors'' in logical
dimension~$\hat{\imath}$, which are the spectral modes marginalized over the two
other logical dimensions~$\hat{\jmath},\hat{k}\neq\hat{\imath}$. They are given by~\cite{Szilagyi2014fna}
\begin{equation}
  P_{q_{\hat{\imath}}} = \sqrt{\frac{N_{k,\hat{\imath}}}{N_k}
    \sum_{p }^{N_k} |\tilde{u}_q|^2 \delta_{p_{\hat{\imath}} q_{\hat{\imath}}}}
  \,\text{,}
\end{equation}
where the spectral modes~$\tilde{u}_q=\mathcal{V}^{-1}_{qp} u_p$ are obtained
from the nodal data~$u_p$ using the Vandermonde
matrix~$\mathcal{V}_{pq}=\prod_{\hat{\imath}}
\Phi_{q_{\hat{\imath}}}(\xi_{p_{\hat{\imath}}})$, and
$\Phi_{q_{\hat{\imath}}}(\xi)$ are the normalized Legendre polynomials.
Given the relative truncation error estimate in each dimension, the $p$-AMR
criterion increases the number of grid points in logical
dimension~$\hat{\imath}$ of an element~$\Omega_k$, and hence the polynomial
order of the expansion~\eqref{eq:lagr_expansion}, if any tensor component of the
variables~$u$ satisfies the condition
\begin{equation}\label{eq:p_amr_crit}
  10^{-\mathcal{T}[P_{q_{\hat{\imath}}}]} \, \max(|u_p|) > \mathcal{T}_\text{target}
  \text{,}
\end{equation}
where $\mathcal{T}_\text{target}$ is a given target absolute truncation error.
For further details see \ccite{Szilagyi2014fna}, Section~5.1. We show that this
$p$-AMR criterion works well even on the extremely stretched DG grids, and that
we do not need to use the compactified grid that was necessary in
\ccite{Vincent2019qpd}.

\subsection{Poisson problem with Lorentzian solution}\label{sec:poisson_lorentzian}

First, we test the DG scheme on a simple Poisson problem with a solution that
falls off as $1/r$ at large distances. We choose the Lorentzian solution
\begin{equation}\label{eq:lorentzian}
  u_\mathrm{analytic}(\bm{x}) = \Big(1 + r^2\Big)^{-1/2}
  \text{,}
\end{equation}
where $r = \sqrt{x^2 + y^2 + z^2}$ is the coordinate distance from the origin.
To solve for this solution we choose the fixed source $\dgf(\bm{x})=-\nabla^2
u_\mathrm{analytic}(\bm{x})=3\left(1+r^2\right)^{-5/2}$ and solve for
$u(\bm{x})$.

This problem was also solved in \ccite{Vincent2019qpd}. In fact, the generic
weak-primal DG scheme developed here reduces to the scheme used in
\ccite{Vincent2019qpd} for a flat-space Poisson equation, except for the
dealiasing techniques employed in \ccite{Vincent2019qpd} that we have not found
to be necessary so far, and some small differences in the definition of the
penalty factor~$\sigma$.

\begin{figure}
  \centering
  \includegraphics[width=\columnwidth]{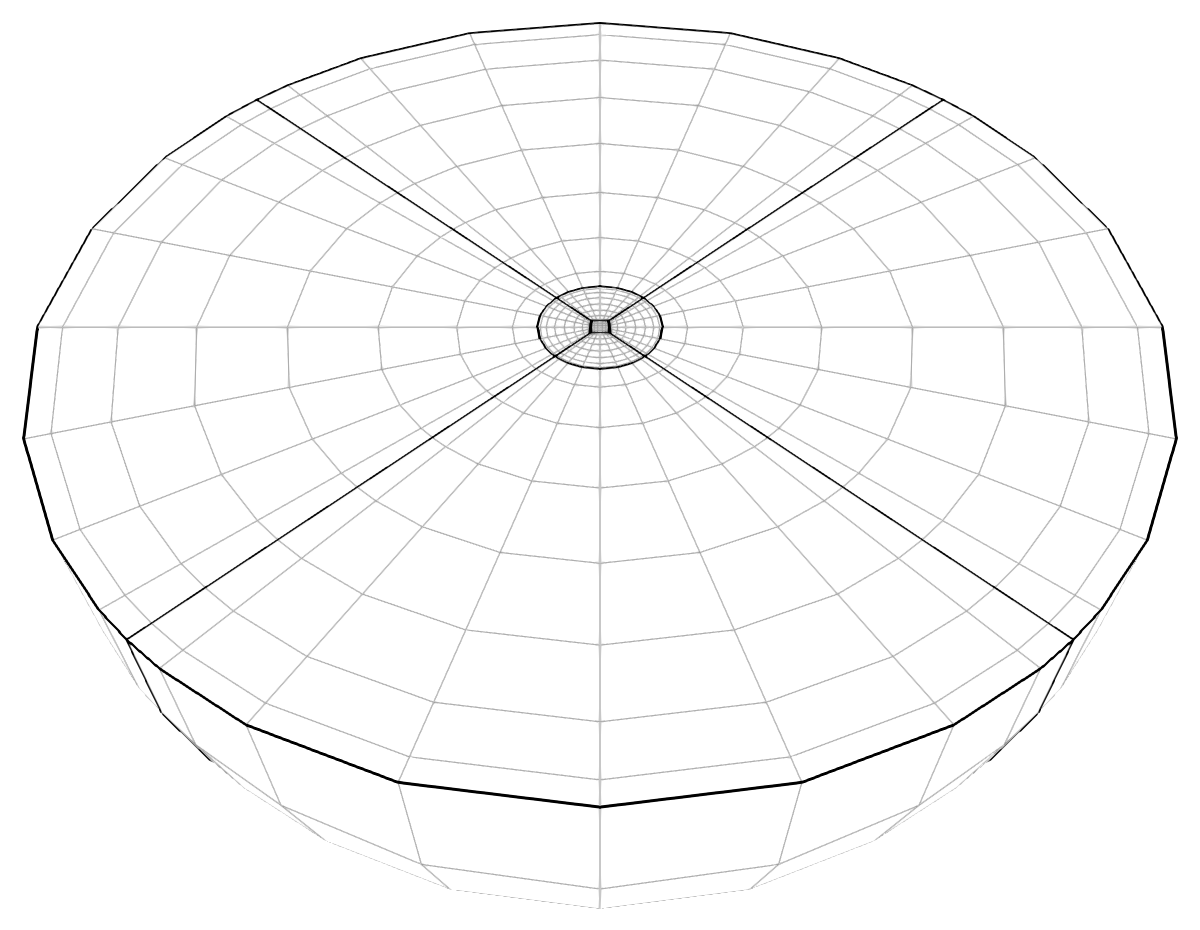}
  \caption{
    \label{fig:lorentzian_domain}
    Spherical domain used in the Lorentzian problem,
    \cref{sec:poisson_lorentzian}. Two spherical shells wrap a central cube.
    Each shell is composed of six wedge-shaped elements (black lines) with a
    grid of LGL collocation points in the element (gray lines). The inner wedges
    transition from a cubical inner boundary to a spherical outer boundary with
    a linear radial mapping. The outer wedges stretch to $R=10^9$ with an
    inverse radial mapping. For this visualization, the inverse radial mapping
    in the outer shell is undone to be able to see both shells.
    Note that while the outermost circle of grid points map to $R=10^9$
    (outermost black circle), the next-to-outermost circle of grid points shown
    here already only map to $r = 400$ (outermost gray circle) due to
    the extreme grid stretching. }
\end{figure}

We choose a spherical domain with outer radius $R=10^9$ and impose the solution,
\cref{eq:lorentzian}, as Dirichlet boundary conditions on the outer boundary.
The domain is illustrated in \cref{fig:lorentzian_domain}. It consists of two
spherical shells wrapped around a central cube. The spherical shells are
composed of six wedges, each with a grid of LGL collocation points, as pictured
in \cref{fig:wedge}. LG collocation points work equally well for this problem
(not shown). The wedges in the
inner shell transition from a cubical inner boundary to a spherical outer
boundary with a linear radial mapping, and the wedges in the outer shell employ
an inverse radial mapping to stretch the grid to the outer boundary at radius
$10^9$.

\begin{figure}
  \centering
  \includegraphics[width=\columnwidth]{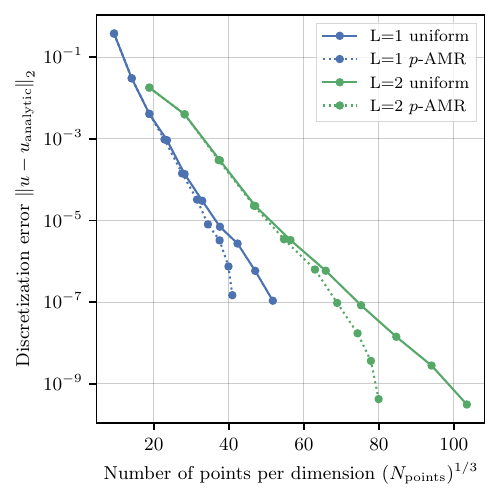}
  \caption{
    \label{fig:lorentzian_convergence}
    Convergence of the DG scheme with increasing polynomial order~$P=1$ to
    $P=10$ for two refinement levels $L$ in the Lorentzian problem,
    \cref{sec:poisson_lorentzian}. The error is measured as an $L_2$~norm over
    all grid points of the difference to the exact solution,
    \cref{eq:lorentzian}. The error decreases exponentially with increasing
    polynomial order~$P$ as expected (solid lines), and the $p$-AMR criterion is
    able to handle the extreme grid stretching as well (dotted lines). }
\end{figure}

\Cref{fig:lorentzian_convergence} shows that the DG scheme achieves the expected
exponential convergence with increasing polynomial order~$P$ for multiple
refinement levels~$L$. The refinement levels are constructed from the domain in
\cref{fig:lorentzian_domain} by splitting each of the 13 elements in two along
each dimension, so at $L=1$ it has 104 elements, and at $L=2$ it has 832
elements. At every refinement level we either increase the polynomial order in
each element and dimension from $P=1$ to $P=10$ uniformly, or we increase the
polynomial order adaptively and anisotropically using the $p$-AMR criterion,
\cref{eq:p_amr_crit}, with $\mathcal{T}_\text{target}=10^{-7}$ for $L=1$ and
$\mathcal{T}_\text{target}=10^{-9}$ for $L=2$. The error converges
exponentially, and the $p$-AMR criterion is able to handle the extreme grid
stretching without the compactified grid that was necessary in
\ccite{Vincent2019qpd}.

\subsection{Puncture initial data with black holes}\label{sec:punctures}

Next, we solve a classic puncture problem for binary black hole initial
data~\cite{Brandt:1997tf}. This method is widely used in the numerical
relativity community to generate initial data because of its simplicity. It is
characterized by the puncture equation,
\begin{equation}\label{eq:puncture}
-\Delta u = \beta \left(\alpha \left(1 + u\right) + 1\right)^{-7}
\text{,}
\end{equation}
where $\Delta$ is the flat-space Laplacian and $u(\bm{x})$ is the puncture field
that we solve for. The background fields $\alpha(\bm{x})$ and $\beta(\bm{x})$
are given by~\cite{Bowen:1980yu}
\begin{subequations}\label{eq:punctures_background}
\begin{align}
&\frac{1}{\alpha} = \sum_I \frac{M_I}{2 r_I}
\text{,}\quad
\beta = \frac{1}{8} \alpha^7 \bar{A}_{ij} \bar{A}^{ij} \\
&\bar{A}^{ij} = \frac{3}{2}\sum_I \frac{1}{r_I^2} \Big(
  \begin{aligned}[t]
  & 2 P_I^{(i} n_I^{j)} - (\delta^{ij} - n_I^i n_I^j) P_I^k n_I^k \\
  & + \frac{4}{r_I} n_I^{(i} \epsilon^{j)kl} S_I^k n_I^l \Big)
\end{aligned}
\end{align}
\end{subequations}
to represent any number of black holes with masses~$M_I$, positions~$\bm{C}_I$,
linear momenta~$\bm{P}_I$, and spins~$\bm{S}_I$. Here, $r_I = \lVert\bm{x} -
\bm{C}_I\rVert$ is the Euclidean coordinate distance to the $I$th black hole and
$\bm{n}_I = (\bm{x} - \bm{C}_I)/r_I$ is the radial unit normal to the $I$th black hole.
Since the puncture equation is just a Poisson equation with a nonlinear source,
it is easily represented by our generic DG scheme.

The puncture equation is an example of a problem with a nonsmooth solution,
which poses a challenge for spectral methods. Specifically, the puncture field
is only $\mathcal{C}_2$ continuous at the punctures~\cite{Brandt:1997tf}, so
spectral methods are not expected to converge exponentially in elements that
contain a puncture. Therefore, the most prominent implementation of the puncture
method in the community, the \texttt{TwoPunctures} code~\cite{Ansorg:2004ds},
uses a special coordinate transformation to achieve rapid (but subexponential)
convergence despite the nonsmooth source. Here we achieve exponential
convergence with our generic DG scheme by an $hp$~refinement strategy, which
presents an alternative and more flexible approach to the problem. We have
solved a similar problem in \ccite{Vincent2019qpd} with a prototype code that
was limited to Poisson-type equations.

\begin{figure}
  \centering
  \tikzinput{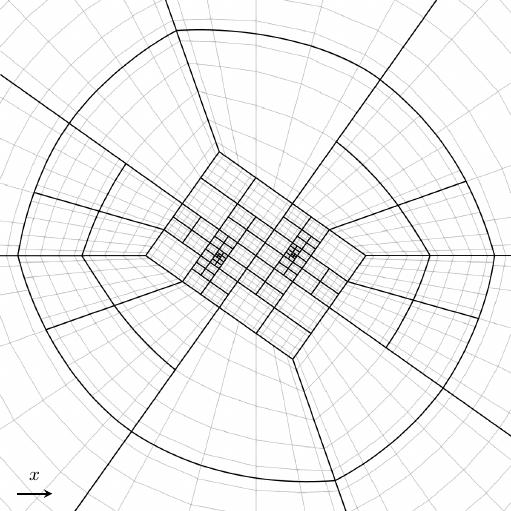}
  \caption{
    \label{fig:punctures_domain}
    Spherical domain used in the puncture problem, \cref{sec:punctures}, after
    six $hp$~refinement steps. The domain is rotated such that the $x$ axis with
    the punctures at $x=\pm 3$ pierces the central cube diagonally. Shown is a
    slice through the domain spanned by the $x$ axis and another diagonal of the
    central cube. }
\end{figure}

We employ the spherical domain already used in the Lorentzian problem,
\cref{sec:poisson_lorentzian}, which is pictured in
\cref{fig:lorentzian_domain}. For the puncture problem we place the outer radius
at only $R=10^5$ and employ Robin boundary conditions,
\begin{equation}
  \partial_r (r \, u) = 0
  \text{.}
\end{equation}
They are implemented as Neumann-type boundary conditions
$(n_i\dgF^i)_\mathrm{N}=-u/r$ in \cref{eq:neumann}.

We can place any number of punctures on the grid to evaluate the background
fields, \cref{eq:punctures_background}. Here we choose the same configuration
that is solved in \ccite{Ansorg:2004ds}, Section~V, which consists of two
orbiting punctures placed at $\bm{x}_\pm=(\pm 3,0,0)$ with masses $M_\pm=0.5$,
linear momenta $\bm{P}_\pm=(0,\pm 0.2,0)$, and spins  $\bm{S}_\pm=0$.
To avoid placing the punctures on element boundaries, we rotate the domain such
that the $x$~axis pierces the central cube diagonally and set the length of the
diagonal to $6\left\|\bm{x}_\pm\right\|=18$ (see \cref{fig:punctures_domain}).
This precaution ensures that the $\mathcal{C}_2$ punctures always lie within
elements, demonstrating that the DG scheme can indeed handle such nonsmooth
solutions. Placing the punctures on element boundaries works equally well, as
long as the punctures are not placed exactly on a grid point where the
background quantities in \cref{eq:punctures_background} diverge.

\begin{figure}
  \centering
  \includegraphics[width=\columnwidth]{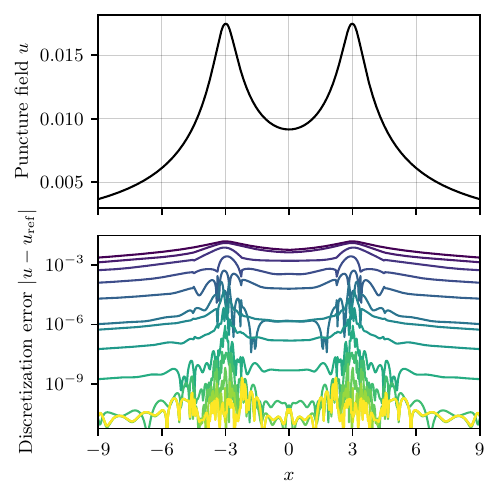}
  \caption{
    \label{fig:punctures_line}
    \emph{Top:} Solution to the puncture problem on the $x$~axis at the highest
    resolution. The punctures are placed at $x=\pm 3$ on this axis. The solution
    is interpolated to the $x$~axis using the Lagrange basis within each
    element, \cref{eq:lagr_expansion}.
    \emph{Bottom:} Numerical error of the solution at increasing resolution. The
    error is computed as the difference to the highest resolution shown in the
    top panel. The error at the punctures is also plotted in
    \cref{fig:punctures_convergence}. }
\end{figure}

\begin{figure}
  \centering
  \includegraphics[width=\columnwidth]{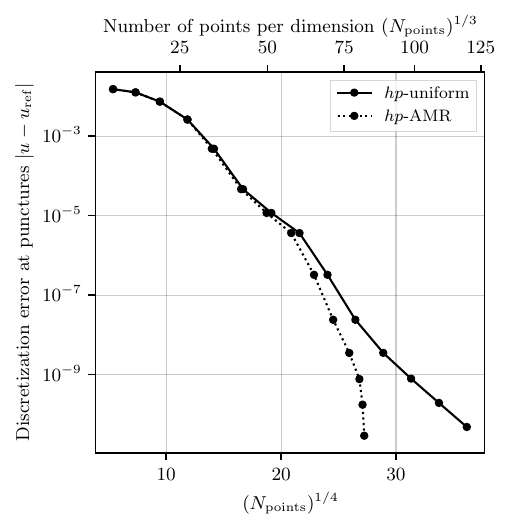}
  \caption{
    \label{fig:punctures_convergence}
    Convergence of the discretization error at the position of the punctures
    ($x=\pm 3$ in \cref{fig:punctures_line}). The error converges exponentially
    with the expected rate of $(N_\text{points})^{1/4}$ (solid line), and the
    $p$-AMR criterion is able to handle the extreme grid stretching as well
    (dotted lines). }
\end{figure}

To demonstrate exponential convergence with our DG scheme we increase the
resolution of the domain iteratively as follows: in every refinement step we
split elements that contain a puncture in two in each dimension
($h$~refinement), and increase the polynomial degree of all other elements by
one in each dimension ($p$~refinement). During this procedure we always impose a
two-to-one balance at element interfaces, meaning that elements may only have
two neighbors per dimension on every side. We enforce the two-to-one balance by
splitting elements at interfaces where the condition is violated. This strategy
leads to the domain pictured in \cref{fig:punctures_domain} after a few
refinement steps, and to the exponential convergence shown in
\cref{fig:punctures_line}. The error at the location of the punctures is also
plotted in \cref{fig:punctures_convergence}. It decreases exponentially with the
expected convergence rate of $(N_\text{points})^{1/4}$~\cite{Schwab2024_privcomm}.
Small variations in the convergence rate are expected because enforcing the
two-to-one balance at element interfaces leads to some h-refinement spilling
into the wedge-shaped elements surrounding the central cube (see
\cref{fig:punctures_domain}).

In addition to the uniform refinement strategy, we also employ the $p$-AMR
criterion, \cref{eq:p_amr_crit}, with $\mathcal{T}_\text{target}=10^{-8}$ to
control the polynomial degree in elements that do not contain a puncture. As in
the Lorentzian problem, \cref{sec:poisson_lorentzian}, we find that the $p$-AMR
criterion performs well on the stretched grids without the compactified grid
that was necessary in \ccite{Vincent2019qpd}. The full $hp$-AMR convergence
study with the 14 refinement levels depicted in \cref{fig:punctures_convergence}
completed in \SI{\sim 7.5}{\minute} on one node of the \texttt{CaltechHPC}
computing cluster at Caltech, running on 56 Intel Xeon Platinum 8276 CPU cores
clocked at \SI{2.2}{\giga\hertz}.

\subsection{Binary black hole XCTS initial data}\label{sec:xcts}

Finally, we test the DG scheme on a binary black hole initial data problem in
the XCTS formulation, \cref{eq:xcts}. The XCTS formulation of the Einstein
constraint equations is more challenging to solve than the puncture formulation
but it has several advantages. For example, it provides a way to impose a
quasiequilibrium condition on the initial data by a choice of $\bar{u}_{ij} = 0$
and $\partial_t K = 0$ and then yields appropriate initial gauge choices for the
evolution in form of the lapse~$\alpha(\bm{x})$ and
shift~$\beta^i(\bm{x})$~\cite{Pfeiffer2004-oo}. The XCTS formulation is also
equipped to generate initial data for black holes with high spins, which is
difficult to achieve with the puncture
method~\cite{Dain:2008ck,Dain:2002ee,Ruchlin:2014zva}. To this end we follow the
\emph{superposed Kerr-Schild} formalism~\cite{Lovelace2008-sw} to set the
conformal metric and the trace of the extrinsic curvature to the superpositions
\begin{subequations}\label{eq:bbh_bg}
\begin{align}
  \bar{\gamma}_{ij} &= \delta_{ij} + \sum_{I=1}^2 e^{-r_I^2 / w_I^2} \, (\gamma_{ij}^I - \delta_{ij})
  \text{,} \\
  K &= \sum_{n=1}^2 e^{-r_I^2 / w_I^2} K^I
  \text{,}
\end{align}
\end{subequations}
where $\gamma_{ij}^I$ and $K^I$ are the spatial metric and the trace of the
extrinsic curvature of two isolated Kerr black holes in Kerr-Schild coordinates,
with $r_I$ the Euclidean coordinate distance from either center. The
superpositions are modulated by two Gaussians with widths $w_I$ so that the
influence of either of the two isolated solutions at the position of the other
is strongly damped. The Gaussians also avoid logarithmic terms in the solution
far away from the center where we employ the inverse radial coordinate mapping,
\cref{eq:invrad}.

Orbital motion is imposed on the two black holes by a split of the shift in a
\emph{background} and an \emph{excess} part~\cite{Pfeiffer2003-Thesis},
\begin{equation}\label{eq:shift_split}
  \beta^i = \beta^i_\bg + \beta^i_\ex
  \text{.}
\end{equation}
We choose the background shift
\begin{equation}\label{eq:bbh_bg_shift}
  \beta^i_\bg = (\bm{\Omega}_0 \times \bm{x})^i + \dot{a}_0 x^i + v^i_0
  \text{,}
\end{equation}
where $\bm{\Omega}_0$ is the orbital angular velocity, $\dot{a}_0$ is a radial
expansion velocity to control the eccentricity of the binary, and $v^i_0$ is a
constant velocity to control its linear
momentum~\cite{Lovelace2008-sw,Ossokine:2015yla}. Then, we insert
\cref{eq:shift_split} into the XCTS equations, \cref{eq:xcts}, and henceforth
solve them for $\beta^i_\ex$ instead of $\beta^i$. This split is
necessary on the stretched grid to avoid numerical error in the derivatives of
the shift far away from the center where $\beta^i_\bg\propto r$.
Instead, derivatives of the background shift are computed analytically and
only derivatives of the variable $\beta^i_\ex$ are computed numerically.

In addition, we find that it is essential on these stretched grids to solve for
$\psi-1$ and $\alpha\psi-1$ instead of $\psi$ and $\alpha\psi$ to avoid
numerical error in the derivatives of the conformal factor and the lapse. This
is because numerical cancellation errors occur in arithmetic operations between
field values at large distances that differ from one only by a small amount.
This effect is further amplified by the conditioning of the dense spectral
differentiation matrices, \cref{eq:diffmat}, which couple field values across
the entire stretched grid of an element.

\begin{figure}
  \centering
  \includegraphics[width=\columnwidth]{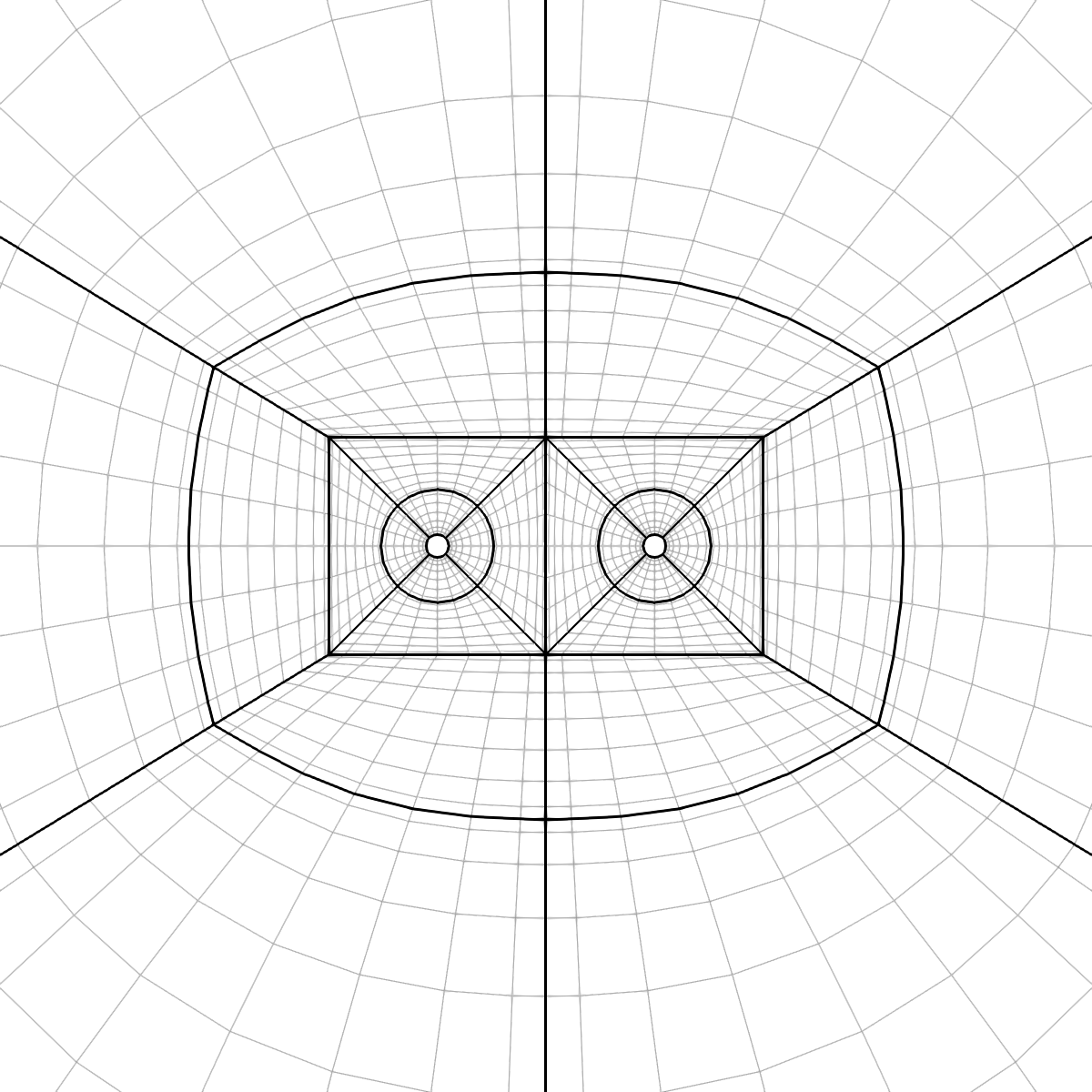}
  \caption{
    \label{fig:bbh_domain}
    Binary black hole domain used in the XCTS initial data problem,
    \cref{sec:xcts}. The domain transitions from two excised spheres in the
    center, to two cubes, to an enveloping sphere, all the way to a spherical
    outer boundary at a radius of up to $R=10^9$ (not shown). The enveloping
    sphere is split in half once in radial direction (black ellipse), so the
    domain shown here has 54 elements in total. }
\end{figure}

Given these modifications to the XCTS formulation we solve the equations on the
domain illustrated in \cref{fig:bbh_domain}. The domain transitions from two
excised spheres that represent the black holes to a spherical outer boundary at
radius~$R$.
At the two excision surfaces in the interior of the domain we impose
apparent-horizon boundary conditions~\cite{Cook2004-yf} with an optional
negative expansion~\cite{Varma2018-fp},
\begin{subequations}\label{eq:ah_bc}
\begin{align}
n^k\bar{\nabla}_k\psi &= \begin{aligned}[t]
  &\frac{\psi^3}{8\alpha}n_i n_j\left(
    (\bar{L}\beta)^{ij} - \bar{u}^{ij}\right) \\
  &- \frac{\psi}{4}\bar{m}^{ij}\bar{\nabla}_i n_j - \frac{1}{6}K\psi^3
  - \frac{\psi^3}{4}\Theta_I
  \text{,}
  \end{aligned} \\
\label{eq:ah_beta}
\beta^i &= \begin{aligned}[t]
  &-\frac{\alpha}{\psi^2} n^i + (\bm{\Omega}^I_r \times (\bm{x} - \bm{x}_I))^i \\
  &+n^i \psi^{-2} (n_j \beta_I^j + \alpha_I)
  \text{,}
  \end{aligned}
\end{align}
\end{subequations}
where $n^i$ is the unit normal to the excision surface pointing out of the
computational domain towards the center~$\bm{x}_I$ of the excision,
$\bar{m}^{ij} = \bar{\gamma}_{ij} - n_i n_j$ is the conformal surface metric,
and $\bm{\Omega}^I_r$ are the rotation parameters that induce spin on the black
hole. The expansion~$\Theta_I$ can be set to zero to impose that the excision
surface \emph{is} an apparent horizon, or to a negative value obtained by
evaluating the isolated Kerr solution at the location of the excision
surface~\cite{Varma2018-fp}. The last term in \cref{eq:ah_beta} also corresponds
to the negative-expansion boundary conditions and is obtained by evaluating the
lapse~$\alpha_I(\bm{x})$ and shift~$\beta_I^j(\bm{x})$ of the isolated Kerr
solution~\cite{Varma2018-fp}. In addition to \cref{eq:ah_bc} we impose the lapse
of the isolated Kerr solution as Dirichlet boundary condition on the excision
surface.\footnote{Note that the \spec{} code typically imposes the
\emph{superposed} lapse on the excision surfaces~\cite{Varma2018-fp}, rather
than the lapse of the isolated Kerr solutions.}

At the outer boundary we impose one of two types of boundary conditions. Either
we impose standard asymptotically flat Dirichlet boundary conditions,
\begin{equation}
  \psi = 1
  \text{,} \quad
  \alpha \psi = 1
  \text{,} \quad
  \beta^i_\ex = 0
  \text{,}
\end{equation}
or we impose new Robin boundary conditions,
\begin{align}
  \partial_r (r \, \psi) &= 0
  \text{,} \quad
  \partial_r (r \, \alpha \psi) = 0
  \text{,} \quad
  \partial_r (r \, \beta^i_\ex) = 0
  \text{.}
\end{align}
Robin-type boundary conditions for the Einstein constraints have been proposed
as early as \ccites{Cook1993,Cook1991}, but not for the more modern XCTS
formulation. The Robin boundary conditions on $\psi$ and $\alpha\psi$ are
straightforwardly implemented as Neumann-type conditions on the fluxes in
\cref{eq:neumann}, $(n_i \bar{\nabla}^i \psi)_\mathrm{N} = -\psi/r$ and $(n_i
\bar{\nabla}^i \alpha\psi)_\mathrm{N} = -\alpha\psi/r$. To implement the Robin
boundary condition for $\beta^i_\ex$ we recast it as a Neumann-type condition on
the flux $n_i\dgF^{\,i}_{\beta_\ex}=n_i(\bar{L}\beta_\ex)^{ij}$ as well. To this
end, we define the projection operator $P^i_j = \delta^i_j - n^i n_j$ to set the
normal component of the shift gradient,
\begin{equation}
  (\bar{\nabla}_i\beta^j_\ex)_\mathrm{N} = P_i^k\bar{\nabla}_k\beta_\ex^j
    -n_i \frac{\beta_\ex^j}{r}
  \text{,}
\end{equation}
and then apply the longitudinal operator, \cref{eq:longitudinal_op}, to obtain
\begin{align}
  (n_i (\bar{L}\beta_\ex)^{ij})_\mathrm{N} &= \begin{aligned}[t]
    &n_i (\bar{L}\beta_\ex)^{ij} - \Big[
    \frac{\beta_\ex^j}{r}
    + n^k \bar{\nabla}_k \beta_\ex^j\Big] \\
    &- \frac{1}{3}n^j n_i\Big[\frac{\beta_\ex^i}{r}
    + n^k \bar{\nabla}_k \beta_\ex^i\Big]
\end{aligned}
\end{align}
We have assumed conformal flatness at the outer boundary in this derivation,
which is given approximately by the exponentially damped superposition in
\cref{eq:bbh_bg}.

\begin{figure}
  \centering
  \includegraphics[width=\columnwidth]{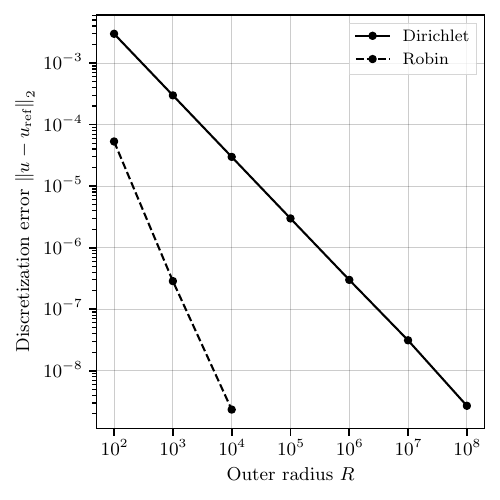}
  \caption{
    \label{fig:bbh_bc}
    Error incurred by the Dirichlet and Robin outer boundary conditions in the
    XCTS problem, \cref{sec:xcts}, with increasing radius of the outer
    boundary~$R$. The Dirichlet boundary conditions incur an error of
    order~$\mathcal{O}(1/R)$, whereas the Robin boundary conditions only incur an
    error of order~$\mathcal{O}(1/R^2)$, thus allowing to place the outer
    boundary closer to the center.}
\end{figure}

We solve for an equal-mass and nonspinning black hole binary in this article.
Our implementation also supports unequal masses and spins, but neither is
particularly relevant to test the extremely stretched grids and outer boundary
conditions. All parameters are listed in the supplementary material. We compute
a discretization error $\left\|u-u_\mathrm{ref}\right\|_2$ as the $L_2$~norm
over all grid points and all variables
$u_\alpha=\{\psi,\alpha\psi,\beta_\ex^i\}$ of the difference to a
high-resolution reference simulation. For the reference simulation we use a
uniform polynomial degree of $P=18$. To evaluate the reference solution at the
grid points of the lower-resolution domain we use the Lagrange interpolating
polynomial basis within the elements, \cref{eq:lagr_expansion}.

\Cref{fig:bbh_bc} compares the two types of boundary conditions at the outer
boundary. The error is measured as the difference to the reference simulation
with outer radius $R=10^9$ and Dirichlet boundary conditions. As expected, the
error incurred by the Dirichlet boundary conditions is of
order~$\mathcal{O}(1/R)$, whereas the error incurred by the Robin boundary
conditions is of order~$\mathcal{O}(1/R^2)$. Therefore, in the following we
place the outer boundary at $R=10^9$ when using Dirichlet boundary conditions,
and at $R=10^5$ when using Robin boundary conditions.

\begin{figure}
  \centering
  \includegraphics[width=\columnwidth]{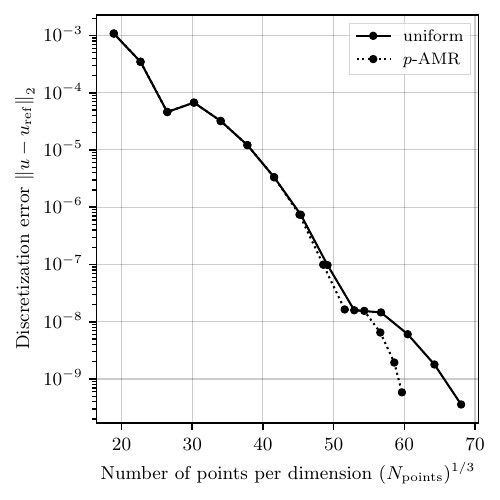}
  \caption{
    \label{fig:bbh_convergence}
    Convergence of the DG scheme with increasing polynomial order $P=4$ to
    $P=17$ for the XCTS problem, \cref{sec:xcts}. The error converges
    exponentially (solid line) and the $p$-AMR criterion is able to handle the
    extreme grid stretching as well (dotted lines). }
\end{figure}

\begin{figure*}
  \centering
  \includegraphics[width=\textwidth]{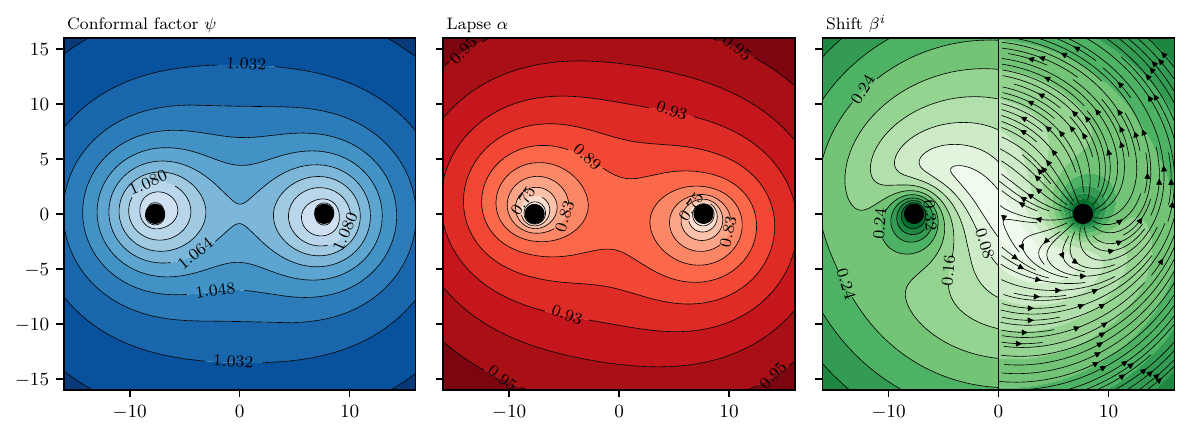}
  \caption{
    \label{fig:bbh_slice}
    Solution to the binary black hole initial data problem in the XCTS
    formulation, \cref{sec:xcts}.}
\end{figure*}

\Cref{fig:bbh_convergence} shows the convergence of the DG scheme for the XCTS
problem with increasing polynomial order~$P$ in each element. We either increase
the polynomial order in all elements uniformly by one in each dimension (solid
line), or apply the $p$-AMR criterion, \cref{eq:p_amr_crit}, with
$\mathcal{T}_\text{target}=10^{-9}$ (dotted line). The discretization error
decreases exponentially on the extremely stretched grids, and the $p$-AMR
criterion is able to handle the extreme grid stretching as well.

The convergence for Dirichlet and Robin boundary conditions in
\cref{fig:bbh_convergence} is identical and not shown. However, the smaller
outer radius for the Robin boundary conditions improves the conditioning of the
discretized problem and hence the convergence rate of the iterative elliptic
solver, reducing its time to solution by about \SI{30}{\percent}. The full
$p$-AMR convergence study with the 14 refinement levels depicted in
\cref{fig:bbh_convergence} completed in \SI{\sim 25}{\minute} on one node of the
\texttt{CaltechHPC} computing cluster at Caltech, running on 56 Intel Xeon
Platinum 8276 CPU cores clocked at \SI{2.2}{\giga\hertz}. This performance
matches the results we have demonstrated in \ccite{ellsolver} despite the
increase in domain radius by many orders of magnitude. Therefore, we are now
able to generate high-accuracy initial data for binary black holes in the XCTS
formulation with the scalable and parallelizable DG method.
\Cref{fig:bbh_slice} shows the solution of the XCTS problem at the highest
resolution.

\section{Conclusion}\label{sec:conclusion}

We have developed a generic discontinuous Galerkin (DG) scheme for elliptic
equations that is able to handle extremely stretched grids for initial data
problems in numerical relativity. Important ingredients for the DG scheme to
handle the extreme grid stretching are: a primal DG formulation, the weak form
of the DG residuals, and the choice to solve for $\psi-1$ and $\alpha\psi-1$
instead of $\psi$ and $\alpha\psi$ for the XCTS equations.

With these ingredients we have demonstrated exponential convergence of the DG
scheme for three test problems with Jacobians as large as $\jac\sim 10^{35}$: a
Poisson problem with a Lorentzian solution, a puncture problem for binary black
hole initial data, and a binary black hole initial data problem in the XCTS
formulation. The scheme achieves exponential convergence with increasing
polynomial order and is able to handle anisotropic adaptive mesh
refinement~(AMR) without the compactified grid that was necessary in
\ccite{Vincent2019qpd}.

The generic primal DG scheme is implemented in the open-source numerical
relativity code \spectre{}~\cite{spectre} and solves the test problems with the
iterative multigrid-Schwarz preconditioned Newton-Krylov elliptic solver
detailed in \ccite{ellsolver}. The new DG scheme is significantly simpler
to extend to new elliptic equations than the formulation used in
\ccite{dgscheme1}, since the concept of auxiliary variables is avoided
altogether. The results in this article are reproducible with the input files
provided in the supplementary material.

Using the DG scheme we are able to solve the puncture problem for binary black
hole initial data with exponential convergence, which is a challenging problem
for spectral methods due to its nonsmooth solution. Our approach provides a
more flexible alternative to apply spectral methods to the puncture problem than
the special coordinate transformation used in the \texttt{TwoPunctures}
code~\cite{Ansorg:2004ds}. For example, we are able to handle randomly placed
punctures on the computational grid with no special treatment in our code.

Most importantly, with the DG scheme laid out in this article we are able to
produce binary black hole initial data in the XCTS formulation extending to
distances of~$10^9$ with a DG method for the first time. We also propose new
Robin boundary conditions for the XCTS equations that allow us to place the
outer boundary at only $10^5$, alleviating some of the grid stretching. These
advancements now allow us to generate high-quality conformally curved initial
data for our numerical relativity simulations of binary black holes with the
\spectre{} code. The DG method allows us to parallelize the initial data solver
on computing clusters, as detailed in \ccite{ellsolver}, so typical
configurations solve in a few minutes on a single node of a computing cluster.
In addition to the equal-mass and nonspinning configuration solved in this
article we are able to generate initial data with unequal masses and high spins
in the superposed Kerr-Schild~(SKS) formalism~\cite{Lovelace2008-sw}, as well as
variations such as superposed Harmonic
Kerr~(SHK)~\cite{Varma2018-fp,Ma:2021can,Cook:1997qc} and spherical
Kerr-Schild~\cite{Chen:2021rtb} initial data. We also support negative-expansion
apparent horizon boundary conditions that avoid constraint-violating
extrapolation into the excision~\cite{Varma2018-fp}. Forthcoming work on the
\spectre{} initial data solver for binary black holes will focus on control
loops for the initial data parameters in the spirit of \ccite{Ossokine:2015yla}.

The results presented in this article are also an important step to support
forthcoming work on initial data in theories beyond general
relativity~\cite{Nee:2024bur}, as well as initial data involving neutron stars.
They lay the foundation for high-quality initial data in the \spectre{} code to
be used in the next generation of numerical relativity simulations. Due to the
open-source nature of the \spectre{} code, the initial data solver is
publicly available to the numerical relativity community and the generated
initial data can be imported into other codes as well.\footnote{Instructions are
provided on \url{https://spectre-code.org} and on request from the author.}

\begin{acknowledgments}
NV thanks Harald Pfeiffer, Saul Teukolsky, and Christoph Schwab for helpful
discussions, as well as the \spectre{} development team for the collaborative
work that enabled this research. Computations were performed with the \spectre{}
code~\cite{spectre} on the CaltechHPC cluster at Caltech. The figures in this
article were produced with \texttt{matplotlib}~\cite{matplotlib1,matplotlib2},
\texttt{TikZ}~\cite{tikz}, and \texttt{ParaView}~\cite{paraview}. This work was
supported in part by the Sherman Fairchild Foundation and by NSF Grants No.\
PHY-2011961, No.\ PHY-2011968, and No.\ OAC-1931266.
\end{acknowledgments}

\appendix

\section{Fluxes and sources for the XCTS equations}\label{sec:xcts_fluxes}

The XCTS equations, \cref{eq:xcts}, can be written in first-order flux form,
\cref{eq:fluxform}. For the Hamiltonian constraint, \cref{eq:xcts_hamiltonian},
the fluxes and sources are
\begin{subequations}
\begin{align}
\dgFi{\psi} &= \bar{\gamma}^{ij} \partial_j \psi
\text{,} \\
\dgS_\psi &= \begin{aligned}[t]
  &-\bar{\Gamma}^i_{ij} \dgFj{\psi} + \frac{1}{8}\psi\bar{R}
  + \frac{1}{12}\psi^5 K^2 \\
  &- \frac{1}{8}\psi^{-7}\bar{A}^2 - 2\pi\psi^5\rho
  \text{,}
  \end{aligned}
\end{align}
for the lapse equation, \cref{eq:xcts_lapse}, the fluxes and sources are
\begin{align}
\dgFi{(\alpha\psi)} &= \bar{\gamma}^{ij} \partial_j \alpha\psi
\text{,} \\
\dgS_{(\alpha\psi)} &= \begin{aligned}[t]
  &-\bar{\Gamma}^i_{ij} \dgFj{(\alpha\psi)}
  + \alpha\psi \Bigg(\frac{7}{8}\psi^{-8} \bar{A}^2
  + \frac{5}{12} \psi^4 K^2 \\
  &+ \frac{1}{8}\bar{R}
  + 2\pi\psi^4\left(\rho + 2S\right) \Bigg)
  - \psi^5\partial_t K \\
  &+ \psi^5\left(\beta^i\bar{D}_i K
  + \beta_\bg^i\bar{D}_i K\right)
  \text{,}
  \end{aligned}
\end{align}
and for the momentum constraint, \cref{eq:xcts_momentum}, the fluxes and sources
are
\begin{align}
\dgF^{ij}_\beta &= \left(\bar{L}\beta\right)^{ij} = \bar{\nabla}^i \beta^j
+ \bar{\nabla}^j \beta^i
- \frac{2}{3} \bar{\gamma}^{ij} \bar{\nabla}_k \beta^k
\text{,} \\
\dgS^i_\beta &= \begin{aligned}[t]
  &-\bar{\Gamma}^j_{jk} \dgF^{ik}_\beta
  - \bar{\Gamma}^i_{jk} \dgF^{jk}_\beta \\
  &+ \frac{2 \alpha\psi}{\psi^{7}} \bar{A}^{ij}
  \left(\frac{\partial_j(\alpha\psi)}{\alpha\psi}
  - 7 \frac{\partial_j \psi}{\psi}\right) \\
  &- \bar{D}_j\left(\left(\bar{L}\beta_\bg\right)^{ij}
  - \bar{u}^{ij}\right) \\
  &+ \frac{4}{3}\frac{\alpha\psi}{\psi}\bar{D}^i K
  + 16\pi\left(\alpha\psi\right)\psi^3 S^i
  \text{,}
  \end{aligned}
\end{align}
where
\begin{align}
\bar{A}^{ij} &= \frac{\psi^7}{2\alpha\psi}\left(
\left(\bar{L}\beta\right)^{ij} +
\left(\bar{L}\beta_\bg\right)^{ij} - \bar{u}^{ij} \right)
\end{align}
and all $f_\alpha=0$.
\end{subequations}

\bibliography{References}

\end{document}